\documentclass[letterpaper, 10 pt, conference]{ieeeconf}  
\IEEEoverridecommandlockouts
\usepackage[utf8]{inputenc}
\usepackage[T1]{fontenc}
\usepackage{amsmath}
\usepackage{amsfonts}
\usepackage{graphicx}
\usepackage{xcolor}
\usepackage{figsize}
\usepackage{upgreek}
\usepackage{amssymb}
\usepackage{algorithmic}
\usepackage[linesnumbered,ruled,vlined]{algorithm2e}
\usepackage{bm}
\usepackage{xargs}
\usepackage{soul}
\usepackage{cancel}
\usepackage{mathtools}
\usepackage{verbatim}
\usepackage{graphics}
\usepackage{epsfig} 
\usepackage{xfrac}
\usepackage{tikz}
\usepackage{accents}
\usepackage[normalem]{ulem}
\usepackage{booktabs}
\usepackage{siunitx}
\usepackage{subcaption}
\graphicspath{{figures/}}

\newcommand{\norm}[1]{\left\lVert#1\right\rVert}
\newcommand {\matr}[2]{\left[\begin{array}{#1}#2\end{array}\right]}

\newcommand{\x}{{\mathbf{x}}}
\newcommand{\y}{{\mathbf{y}}}
\renewcommand{\u}{{\mathbf{u}}}
\newcommand{\w}{{\mathbf{w}}}
\newcommandx{\wb}[2][1=n,2=k]{\w_{#1|#2}}

\newcommand{\bx}{{\x}}
\newcommand{\bu}{{\u}}

\newcommand{\bg}{{{b_g}_{n|k}}}
\newcommand{\DDelta}{{\mathbf{\Delta}}}

\newcommandx{\xb}[2][1=n,2=k]{\x_{#1|#2}}

\newcommandx{\zb}[2][1=n,2=k]{\bar\z_{#1|#2}}
\newcommandx{\ub}[2][1=n,2=k]{\u_{#1|#2}}
\newcommandx{\yb}[2][1=n,2=k]{\bar\y_{#1|#2}}
\newcommandx{\vb}[2][1=n,2=k]{\vv_{#1|#2}}
\newcommandx{\rbx}[2][1=n,2=k]{\r_{#1|#2}^{\x}}
\newcommandx{\rbu}[2][1=n,2=k]{\r_{#1|#2}^{\u}}
\newcommandx{\hb}[3][1=n,2=k,3={}]{h_{#1}^{#3}}
\newcommandx{\gb}[3][1=n,2=k,3={}]{g_{#1|#2}^{#3}}
\newcommandx{\gbar}[3][1=n,2=k,3={}]{\bar{g}_{#1|#2}^{#3}}
\newcommandx{\xT}[2][1=n,2=k]{\mathcal{X}_{#1|#2}}

\newtheorem{Theorem}{Theorem}

\newtheorem{Proposition}{Proposition}

\newtheorem{Assumption}{Assumption}
\newtheorem{Definition}{Definition}
\newtheorem{Remark}{Remark}

\title{\LARGE \bf
Priority-Driven Safe Model Predictive Control Approach to Autonomous Driving Applications}

\author{Francesco Prignoli$^{1}$, Ying Shuai Quan$^{2}$, Mohammad Jeddi$^{1,3}$, Jonas Sj$\ddot{\text{o}}$berg$^{2}$,  and Paolo Falcone$^{1,2}$
\thanks{This work is funded by the Fordonsstrategisk Forskning och Innovation program of Vinnova under the grant 2018-05005.
		}
\thanks{Project co-funded by the European Union – Next Generation Eu - under the National Recovery and Resilience Plan (NRRP), Mission 4 Component 1 Investment 4.1 - Decree No 118 of Italian Ministry of University and Research - Concession Decree No. 2333 of the Italian Ministry of University and Research, Project code D93C23000450005, within the Italian National Program PhD Programme in Autonomous Systems (DAuSy).
		}
\thanks{$^{1}$ Francesco Prignoli, Mohammad Jeddi, and Paolo Falcone are with the Dipartimento di Ingegneria ``Enzo Ferrari'' Universit\`a di Modena e Reggio Emilia, Italy {\tt\footnotesize \{francesco.prignoli, mohammad.jeddi, falcone\}@unimore.it }}
\thanks{$^{2}$ Ying Shuai Quan, Jonas Sj$\ddot{\text{o}}$berg and Paolo Falcone are with the Mechatronics group at the Department of Electrical Engineering, Chalmers University of Technology, Gothenburg, Sweden {\tt\footnotesize \{quany,jonas.sjoberg,paolo.falcone\}@chalmers.se }}
\thanks{$^{3}$ Mohammad Jeddi is also with  the Dept. of Electrical and Information Engineering, Italian National Ph.D. DAUSY, Polytechnic of Bari, Italy {\tt\footnotesize \ m.jeddi@phd.poliba.it}}
}

\begin{document}
\maketitle
\thispagestyle{empty}
\pagestyle{empty}
\begin{abstract}



This paper demonstrates the applicability of the \emph{safe} model predictive control (SMPC) framework to autonomous driving scenarios, focusing on the design of adaptive cruise control (ACC) and automated lane-change systems. Building on the SMPC approach with priority-driven constraint softening---which ensures the satisfaction of \emph{hard} constraints under external disturbances by selectively softening a predefined subset of adjustable constraints---we show how the algorithm dynamically relaxes lower-priority, comfort-related constraints in response to unexpected disturbances while preserving critical safety requirements such as collision avoidance and lane-keeping. 
A learning-based algorithm approximating the time consuming SMPC is introduced to enable real-time execution.
Simulations in real-world driving scenarios subject to unpredicted disturbances confirm that this prioritized softening mechanism consistently upholds stringent safety constraints, underscoring the effectiveness of the proposed method.
\end{abstract}
\section{Introduction}
Model predictive control (MPC) has emerged as a powerful tool in autonomous systems, particularly in safety-critical applications such as autonomous driving, due to its ability to handle system dynamics, preview information, and constraints in a unified optimization framework. 
However, a significant challenge in real-world deployment lies in the fact that not all constraints are known at the design stage. 
To address this, safe model predictive control (SMPC) frameworks have been developed to guarantee constraint satisfaction 
even in the presence of a-priori unknown  constraints, provided certain assumptions hold~\cite{batkovic2022safe, batkovic2023experimental}.
However, in practice, especially in autonomous driving scenarios with a highly dynamic and uncertain environment, these assumptions can be violated due to disturbances, such as unmodeled or rapidly changing maneuvers of other road users (RUs).
The behavior of surrounding RUs—such as human-driven vehicles, cyclists, and pedestrians—is often only partially predictable, rendering nominal prediction models insufficient and leading to potential infeasibility in SMPC. 
%

In such settings, an autonomous vehicle must rapidly adapt its planned trajectory and control actions to ensure safe operation. The planning and control layer, in particular, must be both responsive and flexible to effectively handle non-nominal or emergency scenarios.
To support flexible control under uncertainty, MPC formulations have been extended to incorporate time-varying constraints.~\cite{liu2019recursive}. However, these often rely on worst-case environmental assumptions, resulting in overly conservative control strategies. Alternative formulations introduce slack variables with exact penalty functions~\cite{kerrigan_soft_2000,borrelli_predictive_2017}. 
Nevertheless, assigning appropriate penalty weights becomes impractical as the number of slack variables increases. Moreover, priorities are implicitly encoded via cost weights—typically chosen through trial-and-error—offering limited adaptability in real-time changes~\cite{zheng2011advanced}.
%
%
To resolve this issue,~\cite{kerrigan_designing_2002} introduced a lexicographic MPC framework that enforces constraint priorities via sequential optimization, offering a systematic way to enforce priorities. However, it remains rigid: if a high-priority hard objective cannot be met, all lower-priority ones are disregarded, even if they could be achieved without conflict. This lack of flexibility limits its applicability in real-world, complex traffic environments with parallel or interacting objectives.

Motivated by the need for flexible, reliable control in uncertain and rapidly changing driving conditions, we propose a priority-driven constraint softening framework for SMPC specifically tailored for autonomous driving. 
This work leverages a critical distinction between safety constraints, which must always be satisfied, and comfort constraints, which may be temporarily relaxed. By systematically softening comfort constraints in response to unexpected disturbances, the proposed method ensures controller feasibility without compromising safety.
Our method includes feasibility assessment and real-time soft constraint adaptation, enabling the AV to maintain safety even under emergency disturbances from the environment. A learning-based algorithm is employed to approximate the relationship between external disturbances and the required constraint adaptations. By organizing constraint relaxations according to predefined priorities, the proposed framework guarantees both safety and feasibility while remaining computationally efficient.

The effectiveness of the proposed approach is demonstrated through a set of simulated highway driving scenarios, derived from real-world data, and involving sudden and unforeseen environmental changes. Results show that our framework enables robust and adaptive behavior, making it suitable for real-world deployment in autonomous driving applications.
\subsection{Notation}\label{sec:notation}
Consider a discrete-time linear system described by
	\begin{equation}\label{eq:sys}
	\x_{k+1}=f(\x_k,\u_k),		
	\end{equation}
	where $\x_k\in\mathbb{R}^{n_x}$ and $\u_k\in\mathbb{R}^{n_u}$ are the state and input vectors at time $k$, respectively. 
	The system~\eqref{eq:sys} is subject to the state and input 
    constraints 
    $h(\x,\u)\leq{}0$
	and 
    {the state, input and disturbance constraints 
    $ {g(\x,\u, b_g)\leq{}0}$ in the form of
     {
	\begin{equation}\label{eq:g_constr}
			g(\x_{n|k},\u_{n|k},{b_g}_{n|k}) = C_g\x_{n|k} +D_g\u_{n|k} - {b_g}_{n|k},
	\end{equation}
    with \(h(\x,\u)\in \mathbb R^{n_h}\) and \(g(\x,\u,b_g)\in \mathbb R^{n_g}\).
      In the context of mobile autonomous systems, $g(\cdot,\cdot,\cdot)\le0$ 
    arise from the interactions with the environment and the external \emph{measured} disturbance~${b_g}_{n|k}$ is the predicted motion of the surrounding moving obstacles. 
    }
	The predicted state, input and 
      {disturbance} at time $n$ given the information available at time $k$ are denoted by $\x_{n|k}$, $\u_{n|k}$, and $ {{b_g}_{n|k}}$ respectively. 
	We use the notation $\mathbb{I}_{a}^{b} := \{a,a+1,...,b\}$ to denote a set of integers.
	We denote the Euclidean norm as $\norm{x} = \sqrt{x^T x}$ whereas the Euclidean norm w.r.t. $P=P^T\succ 0$ is denoted by $\| x \|^2_P=x^T P x$.
     {We define $\x^\u_+$ as the next state under control input $\u$: $\x^\u_+:=f(\x,\u)$.}

\section{Background}
We define safety, which is the primary objective of our controller:
	\begin{Definition}[Safety]\label{def:safe}
		A controller $\u_k = \kappa(\x_k)$ is safe for the system (\ref{eq:sys}) in a set $\mathcal{S}\subseteq\mathbb{R}^{n_x}$ if $\forall \, \x\in\mathcal{S}$, the control inputs $\mathbf{U}=\{\kappa(\x_0),...,\kappa(\x_\infty)\}$ and the corresponding state trajectories $\mathbf{X}=\{\x_0,\x_1,...,\x_\infty\}$ are such that $h(\x_k,\u_k)\leq{}0$ and~ {$g(\x_k,\u_k,{b_g}_k)\leq{}0$}, $\forall \, k \geq 0$.
	\end{Definition}

Based on the Safety Definition \ref{def:safe}, we formulate the following SMPC control problem. 
    \begin{subequations}\label{eq:nmpc}
		\begin{align}
		\min_{\substack{\bu}}& \sum_{n=k}^{k+N-1}
		q_\mathbf{r}(\xb,\ub)+p_\mathbf{r}(\xb[k+N])\nonumber\\
		\text{s.t.}\ &\xb[k][k] = \x_{k},&\label{eq:nmpcState} \\
		&\xb[n+1] = f(\xb,\ub), &\hspace{-3em}n\in \mathbb{I}_k^{k+M-1} \label{eq:nmpcDynamics}\\
		&h(\xb,\ub) \leq{} 0, & \hspace{-3em}n\in \mathbb{I}_k^{k+M-1} \label{eq:nmpcInequality_known}\\
		& {g(\xb,\ub,\bg)} \leq{} 0, & \hspace{-3em}n\in \mathbb{I}_k^{k+M-1} \label{eq:nmpcInequality_unknown}\\
		&\xb[k+n] \in\mathcal{X}_\mathbf{r}^\mathrm{s},&\hspace{-3em}n\in \mathbb{I}_{k+N}^{k+M-1} \label{eq:nmpcTerminal}\\
		& {\xb[k+M]\in\mathcal{X}_\mathrm{safe}},\hspace{-10em}&\label{eq:nmpcTerminalSafe}
		\end{align}
	\end{subequations}
    where $k$ is the current  time, $N$ is the prediction horizon, while $M \geq N$ is an extended  prediction horizon. 
	The stage and terminal cost functions $q_\mathbf{r}$ and $p_\mathbf{r}$ penalize the deviations from the reference trajectory $\mathbf{r} = (\mathbf{r}^\x, \mathbf{r}^\u)$. 
	The constraint \eqref{eq:nmpcState} ensures that predictions of the systems dynamics in \eqref{eq:nmpcDynamics} start from the current system state~$\x_{k}$. 
	The constraints \eqref{eq:nmpcTerminal} and \eqref{eq:nmpcTerminalSafe} define a stabilizing and a safe set, respectively. It can be shown that the system
    \begin{equation}\label{eq:sys_cl}
	\x_{k+1}=f(\x_k,\kappa(\x_k)),
	\end{equation}
    with~$\kappa(\x_k)=\ub^*$ and~$\{\ub^*\}_{n=k}^{k+M-1}$ solution of~\eqref{eq:nmpc}, is safe according to Definition~\ref{def:safe}, if the following assumptions hold.

    \begin{Assumption}\label{a:cont}
			The system model $f$ is continuous. 
             {The stage cost $q_\mathbf{r}:\mathbb{R}^{n_x}\times \mathbb{R}^{n_u} \rightarrow\mathbb{R}_{\geq{}0}$, and terminal cost $p_\mathbf{r}:\mathbb{R}^{n_x}\rightarrow\mathbb{R}_{\geq{}0}$, are continuous at the origin and satisfy 
            $q_\mathbf{r}({\mathbf{r}^{\x}_k},\mathbf{r}^{\u}_k)=0$, and $p_\mathbf{r}({\mathbf{r}^{\x}_k})=0$.}
			 Additionally, $q_\mathbf{r}(\bx_k,\bu_k)\geq{}\alpha_1(\|\bx_k-{\mathbf{r}^{\x}_k}\|)$ for all feasible $\x_k$, $\u_k$, and  $p_\mathbf{r}(\bx_N)\leq\alpha_2(\|\bx_N-\mathbf{r}^{\x}_N\|)$, where $\alpha_1$ and $\alpha_2$ are $\mathcal{K}_\infty$-functions.
	\end{Assumption}
    
    \begin{Assumption} \label{a:rec_ref}
			The reference is feasible for the system dynamics, i.e., $\mathbf{r}^\x_{k+1}=f(\mathbf{r}^\x_k,\mathbf{r}^\u_k)$, and satisfies the 
            constraints \eqref{eq:nmpcInequality_known}, i.e., 
             {$h(\mathbf{r}^\x_k,\mathbf{r}^\u_k) \leq{} 0$, for all $k\in\mathbb{I}_0^\infty$.}
	\end{Assumption}
	 {\begin{Assumption} \label{a:terminal}
		There exists a parametric stabilizing terminal set $\mathcal{X}^\mathrm{s}_\mathbf{r}$ and a terminal control law $\kappa^\mathrm{s}_\mathbf{r}(\mathbf{x})$ yielding $p_\mathbf{r}(\x^{\kappa^\mathrm{s}_\mathbf{r}(\mathbf{x})}) - p_\mathbf{r}(\x) \leq{} - q_\mathbf{r}(\x,\kappa^\mathrm{s}_\mathbf{r}(\x)$,  $\x^{\kappa^\mathrm{s}_\mathbf{r}(\mathbf{x})}\in\mathcal{X}^\mathrm{s}_\mathbf{r}$, and 
        $h(\x,\kappa^\mathrm{s}_\mathbf{r}(\x)) \leq{} 0$.
	\end{Assumption}}

	Assumptions \ref{a:cont}-\ref{a:terminal} are commonly used in standard MPC to enforce stability, as noted in \cite{rawlings2017model}. 
	To additionally ensure safety, we introduce the following assumptions:
	\begin{Assumption} \label{a:unknown_constraints}
    The 
    constraint function~$g(\cdot,\cdot,\cdot)$ satisfies
     {
        \begin{align}
			g(\xb,\ub,{b_g}_{n|k+1}) &\leq g(\xb,\ub,\bg),
            \label{eq:assump_unrelaxed}
		\end{align}
        }
    for all $n\geq k$.
	\end{Assumption}
	Assumption~\ref{a:unknown_constraints} requires that the 
     constraints $g(\x_k,\u_k,{b_g}_k)\leq{}0$ are \emph{consistent}, that is, they do not become ``more restrictive'' as the system~\eqref{eq:sys} evolves.   {In the context of autonomous mobile robots, Assumption~\ref{a:unknown_constraints} requires that, at time~$k+1$, a moving obstacle does not get ``much'' closer to the robot than what has been predicted at time~$k$.}
	Finally, the following assumption requires the existence of a safe set, which helps ensuring that the controller's recursive feasibility holds. 
     {\begin{Assumption}\label{a:safe}
		There exists a robust invariant set denoted $\mathcal{X}_\mathrm{safe}\subseteq\mathcal{X}_\mathbf{r}^\mathrm{s}$ such that for all $\x\in \mathcal{X}_\mathrm{safe}$ there exists a safe control set $\mathcal{U}_\mathrm{safe}\subseteq\mathbb{R}^{n_u}$ entailing that  $\x^{\u_\mathrm{safe}}\in\mathcal{X}_\mathrm{safe}$, and $h(\x,\u_\mathrm{safe})\leq{}0$, for all $\u_\mathrm{safe}\in\mathcal{U}_\mathrm{safe}$. 
		Moreover, by construction $g(\x,\u_\mathrm{safe},\bg) \leq 0$ for all $\x\in \mathcal{X}_\mathrm{safe}$ and $\u_\mathrm{safe}\in\mathcal{U}_\mathrm{safe}$.
	\end{Assumption}}

	Based on the assumptions \ref{a:cont}-\ref{a:safe}, the following result guarantees the safety of the controller in (\ref{eq:nmpc}).
	\begin{Proposition}\label{prop:rec_feas}
	 \cite{batkovic2022safe}
		\label{lemma:1}
		Suppose that Assumptions \ref{a:cont}, \ref{a:rec_ref}, \ref{a:terminal}, \ref{a:unknown_constraints}, and \ref{a:safe} hold, and that Problem (\ref{eq:nmpc}) is feasible for the initial state $\x_k$. 
		Then, system (\ref{eq:sys}) in a closed loop with the solution of (\ref{eq:nmpc}) applied in receding horizon is safe (recursively feasible) at all times.
		
		\begin{proof}
			 Readers are referred to \cite{batkovic2022safe}.
		\end{proof}
	\end{Proposition}

\section{{Priority-driven softened Safe MPC}}
The assumptions \ref{a:cont}-\ref{a:safe} enable safety guarantees \cite{batkovic2022safe}. However, Assumption \ref{a:unknown_constraints} can be overly restrictive and often violated in practice. The difference~${\Delta_g}_{n|k+1}$, defined as
\begin{equation}
 {\Delta_g}_{n|k+1} = {b_g}_{n|k} - {b_g}_{n|k+1},
\label{eq:g_update}
\end{equation}
describes the deviation of the disturbance predictions made at two consecutive time instants. Note that,~${\Delta_g}_{n|k+1}>0$ implies the violation of the Assumption~\ref{a:unknown_constraints} and, by Proposition~\ref{prop:rec_feas}, the lack of safety guarantees (recursive feasibility). In the context of mobile autonomous systems,~${\Delta_g}_{n|k+1}$  describes the deviation of the position of an obstacle at time~$n$ predicted at time~$k+1$ from the position at the same time instant predicted at time~$k$ and~${\Delta_g}_{n|k+1}>0$ indicates that the obstacle at time~$k+1$ is closer than what has been predicted at time~$k$.

In this section, we design a learning-based SMPC built upon a constraint relaxation mechanism that reacts to the disturbance $b_g$ in~\eqref{eq:g_constr}. 
We show how the measured disturbance~$b_g$ can be used to drive the relaxation of a set of user-defined constraints, ensuring the persistent satisfaction of a set of \emph{hard} constraints---such as those related to collision avoidance.
The key idea is to maintain safety by selectively relaxing a user-defined set of non-critical constraints (e.g., comfort or efficiency) in response to deviations in the disturbance~$b_g$.

This is achieved by formulating a selectively relaxed SMPC problem:
\begin{subequations}\label{eq:lunmpc}
		\begin{align}
		\min_{\substack{\bu}}& \sum_{n=k}^{k+N-1}
		q_\mathbf{r}(\xb,\ub)+p_\mathbf{r}(\xb[k+N])\nonumber\\
		\text{s.t.}\ &(\ref{eq:nmpcState}), (\ref{eq:nmpcDynamics}),  (\ref{eq:nmpcTerminal}),(\ref{eq:nmpcTerminalSafe})& \\
		&\begin{bmatrix}
            h(\xb,\ub) \\
            g(\xb,\ub,\bg)
        \end{bmatrix} \leq{} E{\DDelta^\star}_{n|k}, & \hspace{-2em}~~n\in \mathbb{I}_k^{k+M-1}. 
		\end{align}
\end{subequations}
Here, $E\in \mathbb{R}^{(n_h+n_g)\times n_\Delta}$ is a user-defined  matrix that selects those constraints that can be relaxed, where $E_{ij} = 1$ if the $i$-th row can be relaxed with $j$-th element of $\DDelta$, and $E_{ij} = 0$ otherwise.   Furthermore,~$\sum_{j=1}^{n_\Delta}~E_{ij}=1$ holds~$\forall~i\in\{1,\ldots,n_h+n_g\}$. The optimal relaxation \( \DDelta^\star \) is obtained by solving:
\begin{subequations}\label{eq:slack_opt}
		\begin{align}
		\min_{\substack{\bu,\DDelta}}& 
		\sum_{n=k}^{k+N-1} \norm{{\DDelta}_{n|k}}^2 \nonumber\\
		\text{s.t.}\ & (\ref{eq:nmpcState}),(\ref{eq:nmpcDynamics}), 
        (\ref{eq:nmpcTerminal}),(\ref{eq:nmpcTerminalSafe})& \\
        &0 \leq {\DDelta}_{n|k} \leq \bar{\DDelta},&\hspace{-1em}n\in \mathbb{I}_k^{k+M-1} \\
		&
        \begin{bmatrix}
            h(\xb,\ub) \\
            g(\xb,\ub,\bg)
        \end{bmatrix}
        \leq{} E{\DDelta}_{n|k},  &\hspace{-1em}n\in \mathbb{I}_k^{k+M-1}  \label{eq:relaxed_h}
		\end{align}
	\end{subequations}
where $\bar{\DDelta}$ represents the maximum permissible relaxation, constrained by physical limitations.
This formulation provides a principled method for softening certain constraints in response to disturbances, 
with the aim of preserving safety.
However, solving (\ref{eq:slack_opt}), can be computational demanding for real-time applications. 
Therefore, in the following subsection, we propose training a neural network (NN) to learn an approximation of the solution of the optimization problem~\eqref{eq:slack_opt}.
\subsection{Neural network (NN) approximation for constraint softening}
We define
\begin{equation}
	\boldsymbol \theta_k = 
	\begin{bmatrix}
		\begin{bmatrix}
		{b_g}_{n|k}
	\end{bmatrix}_{n\in \mathbb{I}_k^{k+M-1}}\\
		\x_k
	\end{bmatrix}, 
	\boldsymbol \eta_k = 
    \begin{bmatrix}
		{\DDelta^\star}_{n|k}
	\end{bmatrix}_{n\in \mathbb{I}_k^{k+M-1}},
    \label{eq:inout_NN}
\end{equation}
and introduce a NN described as 
\begin{equation}\label{eq:NN}
    \begin{split}
        \boldsymbol{\eta} = f(\boldsymbol{\theta}) = \begin{bmatrix} f^1(\boldsymbol{\theta}) & f^2(\boldsymbol{\theta}) & \cdots & f^{M-1}(\boldsymbol{\theta}) \end{bmatrix}^\top
    \end{split}
\end{equation}
where $\eta^j = f^j(\boldsymbol{\theta}), \quad j \in \mathbb{I}_1^{M-1}$.
We make the following assumption:
\begin{Assumption}\label{a:fea_set}
	There exists a nonempty compact set 
	${\Theta} = \{ 
		\boldsymbol{\theta}: (\ref{eq:slack_opt}) \ \text{is feasible}
	 \}.$
\end{Assumption}

Furthermore, we assume the NN output \({\hat\DDelta}_{n|k}\) satisfies:  
\begin{Assumption}\label{a:bound_est_error}
	The approximation error is uniformly bounded element-wise for the trained regressor, i.e.,  
	\begin{equation}
		|{\DDelta^\star}_{n|k} - {\hat\DDelta}_{n|k}|\leq \epsilon \mathbf{1}.
	\end{equation}
\end{Assumption}
Finally, the problem~\eqref{eq:slack_opt} can be reformulated as:
 {\begin{subequations}\label{eq:lunmpc}
		\begin{align}
		\min_{\substack{\bu}}& \sum_{n=k}^{k+N-1}
		q_\mathbf{r}(\xb,\ub)+p_\mathbf{r}(\xb[k+N])\nonumber\\
		\text{s.t.}\ &(\ref{eq:nmpcState}), (\ref{eq:nmpcDynamics}),  (\ref{eq:nmpcTerminal}),(\ref{eq:nmpcTerminalSafe})& \\
		&\begin{bmatrix}
            h(\xb,\ub) \\
            g(\xb,\ub,\bg)
        \end{bmatrix} \leq{} E({\hat\DDelta}_{n|k}+\epsilon \mathbf{1}), & \hspace{0em}n\in \mathbb{I}_k^{k+M-1}. 
		\end{align}
	\end{subequations}}
\subsection{Relaxed consistency condition and feasibility}
When combining the NN~\eqref{eq:NN} with the problem (\ref{eq:lunmpc}), it is crucial to guarantee that
the constraints relaxation term \({\hat\DDelta}_{n|k}\) computed by~\eqref{eq:NN} does not overshoot any physical limits. The \emph{Lipschitz constant} of the NN can be used to fulfill such a requirement.
\begin{Definition}
	A function $f: \mathbb{R}^n \rightarrow \mathbb{R}^m$ is globally Lipschitz continuous if there exists an $L \geq 0$ such that:
\begin{equation}
\| f(x)-f(y) \|
\leq
L\| x-y \| ~~
\forall x,y \in \mathbb{R}^n
\label{eq:lips}
\end{equation}
\end{Definition}
The smallest  $L$  for which condition (\ref{eq:lips}) holds is called the Lipschitz constant  $L^*$, which provides an upper bound on how much the output  $f$  can change when the input varies from  $x$  to  $y$ . 

Limits $\bar\DDelta$ on the constraints relaxation~${\DDelta}_{n|k}$ set an upper bound $\bar\eta^j \in \mathbb{R}_+$ for each $\eta^j$. To force the NN to output a relaxation~$\eta^j\le\bar\eta^j$,
we impose the following constraint on \( L^j \)
\begin{equation}
    L^j \leq \bar L^j, \text{with}~\bar L^j=\frac{\bar\eta^j}{\bar\Delta_g+\bar\Delta_\x},
    \label{eq:Lj_bound}
\end{equation}
where $\bar\Delta_g$ specifies the maximum disturbance~$b_g$ 
and $\bar\Delta_\x = \max_{\substack{\x\in \mathrm{Proj}_\x{\Theta}, \u\in\mathcal{U}}} \norm{\x^\u-\x}$.
By enforcing the constraint~\eqref{eq:Lj_bound}, we guarantee that
when 
$\norm{
				\begin{bmatrix}
		{\Delta_g}_{n|k+1}
	\end{bmatrix}_{n\in \mathbb{I}_k^{k+M-1}}}\leq\bar\Delta_g$, then the inequality
\begin{equation*}\label{eq:eta_bound}
    \begin{split}
        \eta^j \leq 
        L^j
			\norm{
				\begin{bmatrix} 
				\begin{bmatrix}
		{\Delta_g}_{n|k+1}
	\end{bmatrix}_{n\in \mathbb{I}_k^{k+M-1}} \\
				\x_{k+1} - \x_{k}
			    \end{bmatrix}}
			\leq
            L^j (\bar\Delta_g+\bar\Delta_\x)
            \leq \bar\eta^j,
    \end{split}
\end{equation*}
holds, that is, the NN outputs a slack variable~$\DDelta\le\bar\DDelta$.
The inequality (\ref{eq:Lj_bound}) can be verified after the NN training phase or enforced during its training using the method proposed in \cite{pauli2021training}. With $L^j$, we then introduce the following assumption, which serves as a relaxed form of Assumption \ref{a:unknown_constraints}:
\begin{Assumption}\label{a:unknown_constraints_relaxed}
		The 
        hard constraint function satisfies
        \begin{align}
            g(\xb,\ub,{b_g}_{n|k+1}) &= g(\xb,\ub,\bg)+{\Delta_g}_{n|k+1},
        \end{align}
        where \begin{align}\label{eq:relax_g}\norm{
				\begin{bmatrix}
		{\Delta_g}_{n|k+1}
	\end{bmatrix}_{n\in \mathbb{I}_k^{k+M-1}}}\leq{\bar\Delta_g}_{k+1},\end{align}
        with
            ${\bar\Delta_g}_{k+1} = \min_j \frac{\bar\eta^j-\epsilon}{\bar L^j}-\norm{\x_{k+1}-\x_k}$.
\end{Assumption}

Given the constraint relaxation bound $\bar\DDelta$, we redefine safety as follows:
 {\begin{Definition}[Safety]\label{def:safe_2}
	A controller $\u_k = \kappa(\x_k)$ is safe for the system (\ref{eq:sys}) in a set $\mathcal{S}\subseteq\mathbb{R}^{n_x}$ if $\forall \, \x\in\mathcal{S}$, the control inputs $\mathbf{U}=\{\kappa(\x_0),...,\kappa(\x_\infty)\}$ and the corresponding state trajectories $\mathbf{X}=\{\x_0,\x_1,...,\x_\infty\}$ are such that $\begin{bmatrix}
            h(\x_k,\u_k) \\
            g(\x_k,\u_k,{b_g}_k)
        \end{bmatrix}
        \leq{} E{\bar\DDelta}$, $\forall \, k \geq 0$.
\end{Definition}}

Now, we present the following result, which ensures safety under a relaxed condition in the presence of external disturbances:
\begin{Theorem}\label{theorem:2} Suppose that Assumptions \ref{a:cont}-\ref{a:terminal}, and \ref{a:safe}-\ref{a:unknown_constraints_relaxed} hold, and that Problem~\eqref{eq:nmpc} is feasible for the initial state $\x_k$. Then, system \eqref{eq:sys} in closed loop with the solution of~\eqref{eq:lunmpc} applied in receding horizon is safe at all times.

\begin{proof}
	Readers are referred to \cite{batkovic2022safe}.
\end{proof}
\end{Theorem}

\subsection{{Priority-driven soft-constrained MPC}}
In the framework proposed in this paper, the designer is asked to define as many matrices~$E^i$ in~\eqref{eq:relaxed_h} as the combinations of constraints that could be possibly relaxed.

Let \( E^i \), \( i \in \mathbb{I}^d_1 \), denote the \( i \)-th \emph{relaxation mode}, ranked from \( 1 \) (lowest priority) to \( d \) (highest priority). 
Thus, as \(i\) increases, higher priority constraints are progressively relaxed.
For each \( E^i \), two NNs are trained: the first, $\text{NN}^1_{E^i}$, is defined as in (\ref{eq:NN}) and trained with \( \boldsymbol \theta \in  \Theta \), while the second, $\text{NN}^2_{E^i}$, outputs an infeasibility indicator, predicting whether the problem~\eqref{eq:slack_opt} is feasible under \( E^i \) and is trained with both \( \boldsymbol \theta \in \Theta \) and \( \boldsymbol \theta \notin \Theta \). Denoting the output of $\text{NN}^2_{E^i}$ as \( \mathcal{F}^i \), we define:
 {\begin{equation}
    \mathcal{F}^i = 
    \begin{cases}
        0, & \text{if (\ref{eq:slack_opt}) with \( E = E^i \) is feasible,} \\
        1, & \text{if (\ref{eq:slack_opt}) with \( E = E^i \) is infeasible.}
    \end{cases}
\end{equation}}

We introduce Algorithm \ref{alg:1} for the proposed priority-driven soft-constrained MPC.
\begin{algorithm}[t]
    \caption{Priority-driven Soft-constrained MPC}
    \label{alg:1}
    \textbf{Initialize:} Set \( \x_0 \).\\
    \For{each time step \( k \)}{
        Obtain \( \x_k \), evaluate \( g(\xb, \ub, {b_g}_{n|k}) \).\\
        \uIf{(\ref{eq:assump_unrelaxed}) holds}{ Solve (\ref{eq:nmpc}). }
        \ElseIf{(\ref{eq:relax_g}) holds}{  
            Compute \( \mathcal{F}^{i} \) via \( \text{NN}^2_{E^i} \) for \( i \in \mathbb{I}^d_1 \).  \\
            \For{lowest-priority \( i \) with \( \mathcal{F}^{i} = 0 \)}{  
                Solve (\ref{eq:lunmpc}) with \( E = E^i \), \( \hat\DDelta \) from \( \text{NN}^1_{E^i} \).  
            }
        }
        \Else{ Report failure, exit. }
    }
\end{algorithm}


\section{Problem formulation}
\subsection{Vehicle Model and System Constraints}
Given a user-defined reference $\mathbf{r}$, we express the vehicle dynamics in the frame of the reference path~
	\begin{equation}
		\begin{split}
			\matr{c}{
			\dot s \\
			\dot e_y \\
			\dot e_\psi \\
			\dot \delta \\
			\dot \alpha \\
			\dot v \\
			\dot a } =
		\matr{c}{
			v \cos(\psi)(1-\kappa^{\mathbf{r}}(s)e_y)^{-1} \\
			v \sin(e_\psi) \\
			\frac{v}{l} \tan(\delta) -\frac{\dot s}{l} \tan(\delta^{\mathbf{r}}(s))\\
			\alpha \\
			w^2_0(\delta^\mathrm{sp}-\delta)-2w_0w_1\alpha \\
			a \\
			t_\mathrm{acc}(a^\mathrm{req}-a)},\\
			\x = 
			\begin{bmatrix}
                s & 
				e_y &
				e_\psi &
				\delta &
				\alpha &
				v &
				a
			\end{bmatrix}^\top,
			\u = 
			\begin{bmatrix}
				\delta^\mathrm{sp} &
				a^\mathrm{req}
			\end{bmatrix}^\top,
		\end{split}
		\label{eq:vehicle_model}
	\end{equation}
	where $s$ represents the longitudinal position along the path,  $v$ is the longitudinal velocity, $\psi$ is the orientation angle in a a global frame, $\kappa^{\mathbf{r}}(s)$ is the path curvature, $e_y$ is the lateral displacement error, $e_\psi$ is the orientation error relative to the reference $\mathbf{r}$,  $l$ is the distance between the front and rear axles, $\delta$ is the steering angle, $\delta^{\mathbf{r}}$ is the reference steering angle, $\alpha$ is the steering angle rate, and $a$ is the acceleration. The control inputs consist of the requested acceleration $a^\mathrm{req}$ and the steering angle set point $\delta^\mathrm{sp}$. Additionally, $w_0$, $w_1$, and $t_\mathrm{acc}$ are model constants that define the steering actuator and acceleration dynamics, respectively.

 \subsection{Cost and system constraints}
   The costs $q_{\mathbf{r}}$ and $p_{\mathbf{r}}$ are defined as quadratic functions 
 of the state and input deviations from the corresponding references~$\mathbf{r}^\x,~\mathbf{r}^\u$, respectively. 
   \begin{equation}
   	\begin{split}
   		q_{\mathbf{r}}(\x_k,\u_k)&:=
   		\matr{c}{
   		\x_k - \mathbf{r}^\x_k\\
   		\u_k - \mathbf{r}^\u_k
   		}^\top W
   		\matr{c}{
   		\x_k - \mathbf{r}^\x_k\\
   		\u_k - \mathbf{r}^\u_k
   		}\\
   		p_{\mathbf{r}}(\x_k)&:=
   		(\x_k - \mathbf{r}^\x_k)^\top P
   		(\x_k - \mathbf{r}^\x_k)
   	\end{split}
   \end{equation}
   where $W:=\mathrm{blockdiag}(Q,R)$ is constructed using positive-definite matrices $Q\in\mathbb{R}^{n_\x\times n_\x}$ and $R\in\mathbb{R}^{n_\u\times n_\u}$. We derive the terminal cost matrix $P\in\mathbb{R}^{n_\x\times n_\x}$ by decoupling the longitudinal and lateral dynamics, as detailed in  \cite{batkovic2023experimental}.

 We assume that system (\ref{eq:vehicle_model}) is subject to the following state and input constraints:
   \begin{gather}\label{eq:box}
   	\norm{e_y}\leq \bar e_y, 
   	\norm{e_\psi}\leq \bar e_\psi, 
   	\norm{\delta}\leq \bar \delta, 
   	\norm{\delta^\mathrm{sp}}\leq \bar \delta,\nonumber\\
   	0\leq v \leq \bar v,   
   	\underline{a} \leq a \leq \bar a, 
   	\underline{a} \leq a^\mathrm{req} \leq \bar a, 
   	\norm{\alpha} \leq \bar \alpha,\\
    \underline{a}_y \leq a_y \leq \bar a_y,
    \underline{j}_y \leq j_y \leq \bar j_y,\nonumber
   \end{gather}
where lateral acceleration and jerk are defined as \(a_y = \dfrac{v^2}{l}\tan(\delta)\) and \(j_y = \dfrac{v^2}{l}\alpha (1+\tan^2(\delta))\), respectively.
Additionally, we define a collision avoidance hard constraint, $g^{\mathrm{safe}}_{n|k}$.
For simplicity, we consider only a single road user (RU), though the approach extends to multiple RUs, as discussed in \cite{batkovic2023experimental}. We introduce a RU's state $\w:=[w^{\mathrm{lon}}_k \ w^{\mathrm{lat}}_k]\in\mathbb{R}^2$ describing the RU's longitudinal and lateral positions along the reference path $\mathbf{r}$. We assume that the uncertainty of the predicted position of the RU is bounded at every time instant, i.e., $\wb \in \mathcal{W}_{n|k} \subseteq \mathbb{R}^{n_\w}$ for all $k$, where $\mathcal{W}_{n|k}$ is an outer approximation obtained via motion prediction and reachability analysis \cite{batkovic2023experimental}.

Given the set $\mathcal{W}_{n|k}$, our objective is to force the ego vehicle avoiding those positions that may result in collisions. We define a safety distance $d_\mathrm{safe}$, accounting for the dimensions of both the ego vehicle and the obstacles. 
The longitudinal positions where the  ego vehicle might collide with the RU are described by the following set:
   \begin{equation}
   	\begin{split}
   		\mathcal{T}_\mathbf{r}(\mathcal{W}_{n|k}):=\{ s \mid \exists & \ \w_{n|k} \in \mathcal{W}_{n|k} ~\mathrm{s.t.} \\&\norm{T^\w(\w_{n|k})- T^\x(\mathbf{r}^\x(s))}_2 \leq d_\mathrm{safe} \}
   	\end{split} 
   	\label{eq:safe_set}
   \end{equation}
   where $T^\x : \mathbb{R}^{n_\x} \rightarrow \mathbb{R}^2 $ and $T^\w : \mathbb{R}^{n_\w} \rightarrow \mathbb{R}^2 $ map the reference position $ \mathbf{r}^\x$ and the uncertain position $\wb$ into a global frame. 
Then defining
\begin{equation}
    \sigma^L_{n|k} := \min_s ~ s, \quad \text{s.t.}~ s \in \mathcal{T}_\mathbf{r}(\mathcal{W}_{n|k}),
\end{equation}
we formulate the 
hard collision avoidance constraint for the yielding strategy, where the ego vehicle stays behind the RU, as
\begin{equation}
    g^{\mathrm{lon}}_{\mathrm{safe}}(s_{n|k}, \sigma^L_{n|k}) := s_{n|k} - \sigma^L_{n|k}.
    \label{eq:g_long_safe}
\end{equation}


To enable a lane change in response to the behavior of the surrounding RUs, the original lateral deviation constraint $\lVert e_y \rVert \leq \bar{e}_y$ is replaced by the following constraint:
\begin{equation}
    g^{\mathrm{lat}}_{\mathrm{safe}}(e_{y,n|k}, \beta^{L}_{n|k}, \beta^{U}_{n|k}) := 
    \begin{bmatrix}
        1 \\ -1 
    \end{bmatrix} e_{y,n|k}
    \leq 
    \begin{bmatrix}
        \beta^{U}_{n|k} \\
        \beta^{L}_{n|k}
    \end{bmatrix},
    \label{eq:g_lat_safe}
\end{equation}
where the lower and upper bounds $\beta^{L}_{n|k}$ and $\beta^{U}_{n|k}$ are defined as
\begin{equation}
\begin{split}
    \begin{cases}
        \beta^{L}_{n|k}(s) = \frac{1}{2}l_w,\quad \beta^{U}_{n|k}(s) = \frac{3}{2}l_w, & 
        \text{left lane}, \\
        \beta^{L}_{n|k}(s) = -\frac{3}{2}l_w,\quad \beta^{U}_{n|k}(s) = -\frac{1}{2}l_w, & \text{right lane},
    \end{cases}\\
    \forall s \in \mathcal{T}_{\mathbf{r}}(\mathcal{W}_{n|k}),
\end{split}
\label{eq:lane bounds}
\end{equation}
with $l_w$ denoting the lane width.
}
In the formulation of the SMPC controller (\ref{eq:nmpc}), the collision avoidance constraints are imposed at those time instants~$n$ when $\mathcal{T}^\mathrm{lon}_\mathbf{r} \neq \emptyset$. We denote such time instants as $n \in \mathbb{I}^{k+l}_{k+m}$.
Note that collision avoidance requires that at least one of \eqref{eq:g_long_safe} or \eqref{eq:g_lat_safe} be satisfied.

In the car-following scenarios, we also define a soft constraint that promote a minimum time headway between the ego vehicle and the RU ahead, formulated as:
\begin{equation}
g_\mathrm{follow}(s_{n|k},\u_{n|k},\sigma^{L}_{n|k}):=
s_{n|k} + t_{\text{gap}} \cdot v_{n|k} - \sigma^{L}_{n|k}.
\label{eq:g_follow}
\end{equation}
where \( t_{\text{gap}} \) is the desired time headway.
This constraint enforces that the ego vehicle maintains a minimum distance that grows proportionally with its speed, thereby accounting for safe braking and reaction time. To accommodate potential violations arising from sudden interactions with surrounding RUs, it is treated as a soft constraint and can be relaxed to prevent unnecessary abrupt braking and to maintain overall feasibility.

\section{Simulation Results}

We evaluate the proposed priority-driven SMPC framework in two highway scenarios involving emergency maneuvers caused by sudden lane changes of surrounding vehicles. Both scenarios are constructed using real-world data from the ZOD dataset~\cite{alibeigi2023zenseact}. Simulations are performed in MATLAB using the CasADi framework~\cite{andersson2019casadi} and IPOPT~\cite{wachter2006implementation}, with a sampling time of \( t_s = 0.1\,\si{\second} \) to match the LiDAR rate used in the dataset. State estimation for surrounding vehicles is performed using a bounded-error observer~\cite{jeon2019tracking}, and the MPC prediction horizon is set to \( N = 20 \), with a constraint learning horizon of \( M = 100 \).
Constant move blocking is applied to the slack variables over the prediction horizon, to reduce the number of decision variables and facilitate the NN training, while still allowing the controller to handle constraint violations.

\vspace{1em}
\subsection{Scenario 1: Adaptive Cruise Control with Constraint Relaxation}

\begin{figure}[t]
\SetFigLayout{3}{1}
\centering
\subfigure{\includegraphics[width=1\columnwidth]{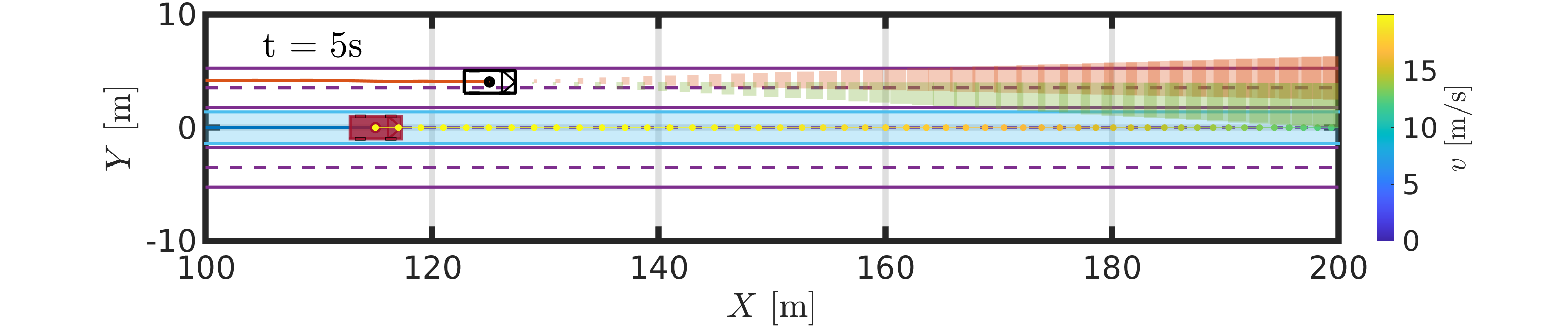}}
\hfill
\centering
\subfigure{\includegraphics[width=1\columnwidth]{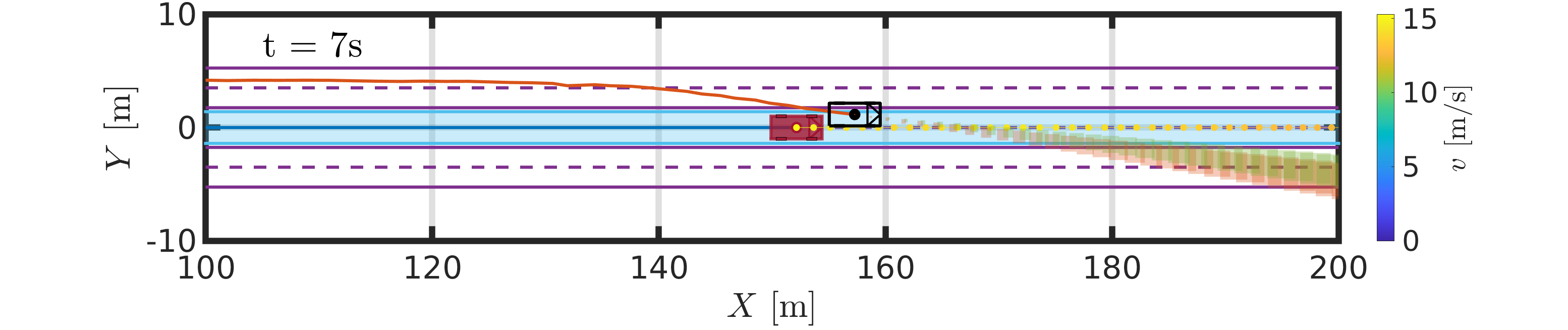}}
\hfill
\centering
\subfigure{\includegraphics[width=1\columnwidth]{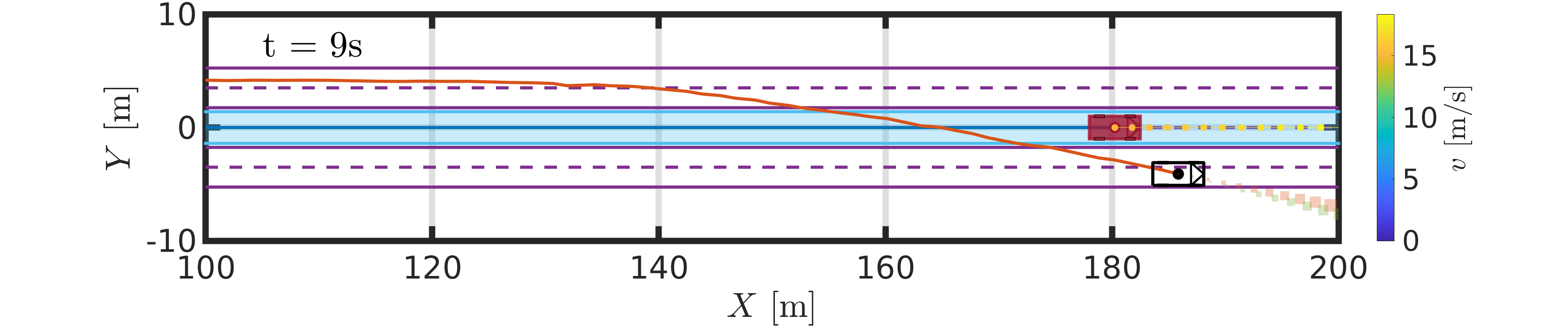}}
\caption{Time sequence in Scenario 1: RU (black) lane crossing forces the ego vehicle (red) to exceed comfort acceleration limits.}
\label{fig:ACC_time_frames}
\end{figure}

\begin{figure}[t]
\SetFigLayout{3}{1}
\centering
\subfigure{\includegraphics[width=0.85\columnwidth]{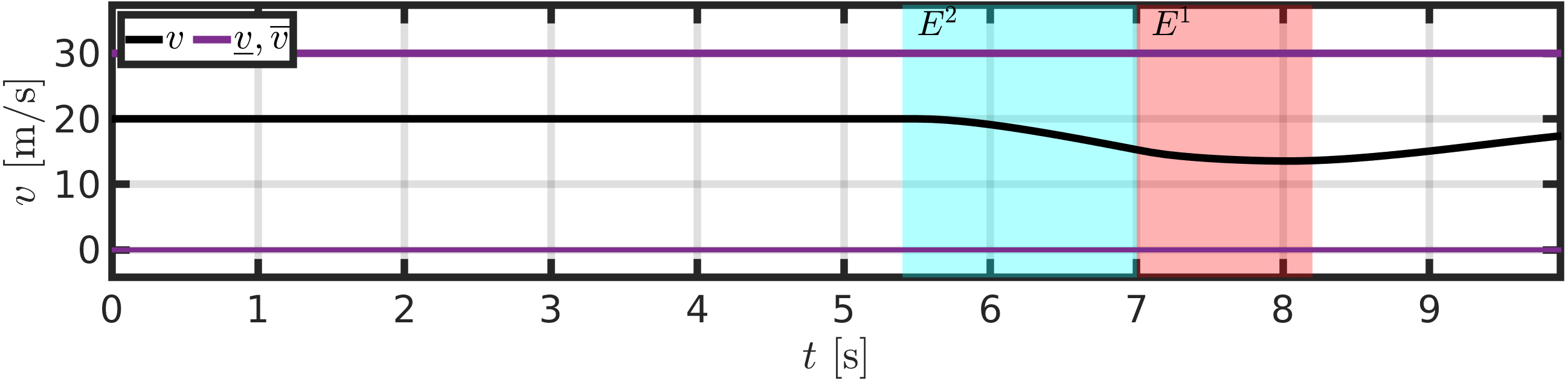}}
\hfill
\centering
\subfigure{\includegraphics[width=0.85\columnwidth]{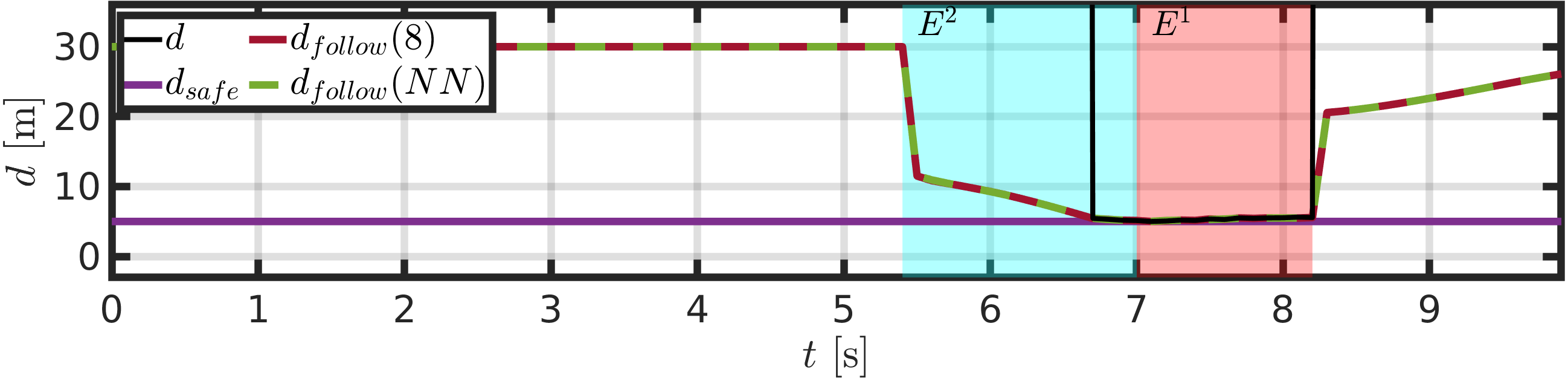}}
\hfill
\centering
\subfigure{\includegraphics[width=0.85\columnwidth]{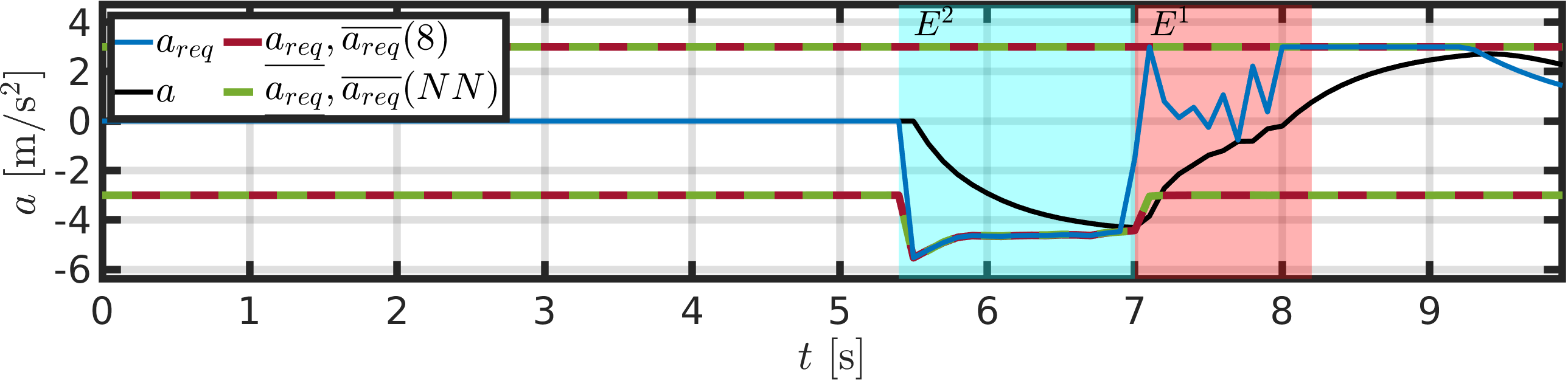}}
\caption{Closed-loop trajectories for Scenario 1: speed (top), distance to the RU (center), and acceleration (bottom).}
\label{fig:ACC_traj_ZOD}
\end{figure}

In the first scenario, a surrounding vehicle suddenly moves from the left to the right lane, cutting across the ego vehicle’s path and violating Assumption~\ref{a:unknown_constraints} (Fig.~\ref{fig:ACC_time_frames}). As illustrated in Fig.~\ref{fig:ACC_traj_ZOD}, the reachable set \( \mathcal{W}_{n|k} \) (orange shaded region) does not contain the subsequent set \( \mathcal{W}_{n|k+1} \) (green shaded region), indicating that Assumption~\ref{a:unknown_constraints} no longer holds.

Here, we consider two relaxation modes, $E^1$ and $E^2$. For each mode \( E^i \), two neural networks \( \text{NN}^1_{E^i} \) and \( \text{NN}^2_{E^i} \) are trained. Each network receives as input the state vector:
\[
\boldsymbol{\theta} = \begin{bmatrix}
\boldsymbol{\sigma}^L & s & v & a
\end{bmatrix}^\top,
\]
where \(\boldsymbol{\sigma}^L=\begin{bmatrix}
    \sigma^L_{n|k}
\end{bmatrix}_{n\in\mathbb{I}^{k+l}_{k+m}}\).
The output slack values of  \( \text{NN}^1_{E^i} \) depend on the relaxation mode.  
In mode \( E^1 \), the network relaxes only the following distance constraint \( g_\mathrm{follow} \), with output \( \eta = \delta_g \).
In mode \( E^2 \), both \( g_\mathrm{follow} \) and the lower bound on the requested acceleration \( \underline{a}^\mathrm{req} \) are relaxed, with output \( \eta = \begin{bmatrix} \delta_a & \delta_g \end{bmatrix}^\top \). 


The activation sequence of each relaxation mode is fully determined by the feasibility indicators, which are outputs of \( \text{NN}^2_{E^i} \).
In Fig.~\ref{fig:ACC_traj_ZOD}, the cyan and red shaded areas show the activation sequence of the two relaxation modes. In the early stage of the maneuver, the relaxation of the following distance alone (mode \(E^1\)) is insufficient, and softening the requested deceleration bound is necessary to avoid a collision (mode \(E^2\)). 
Once the collision is avoided, the controller transitions to mode \(E^1\), where it restores the desired time headway by applying an acceleration that remains within comfortable limits.

\vspace{1em}
\subsection{Scenario 2: Emergency Lane Change Maneuver}
\begin{figure}[t]
\SetFigLayout{3}{1}
\centering
\subfigure{\includegraphics[width=1\columnwidth]{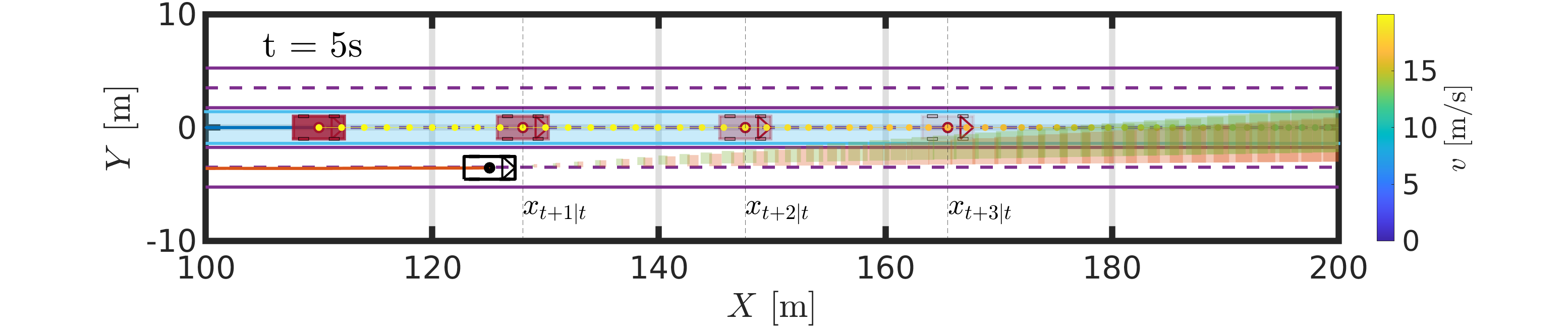}}
\hfill
\centering
\subfigure{\includegraphics[width=1\columnwidth]{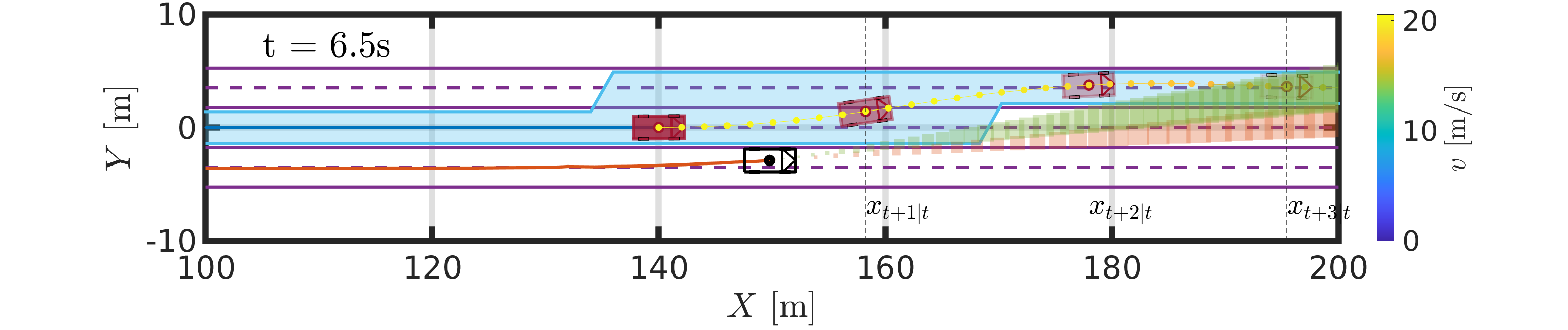}}
\hfill
\centering
\subfigure{\includegraphics[width=1\columnwidth]{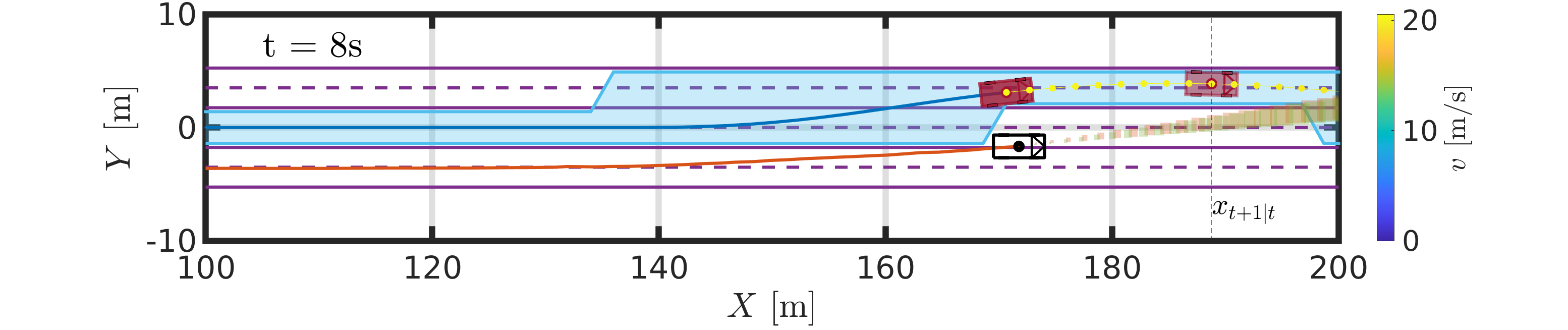}}
\caption{Time sequence in Scenario 2: RU (black) sudden cut-in results in an emergency lane change by the ego vehicle (red).}
\label{fig:LC_time_frames}
\end{figure}

In the second scenario, a surrounding vehicle abruptly cuts into the ego vehicle's lane, creating an imminent collision risk.
In situations where avoiding a longitudinal collision would require hard braking that could exceed the vehicle's braking capability, an evasive lane change maneuver may be the only viable option to preserve safety (Fig. ~\ref{fig:LC_time_frames}).

Here, we consider a new relaxation mode \(E^3\), that softens the lateral acceleration \( a_y \) and jerk \( j_y \) comfort bounds to satisfy lateral safety constraints \eqref{eq:g_lat_safe}, while dropping the longitudinal constraints \eqref{eq:g_long_safe} and \eqref{eq:g_follow}. The two associated neural networks, \(NN_{E^3}^1\) and \(NN_{E^3}^2\), take as input the state vector:
\[
\boldsymbol{\theta} = \begin{bmatrix}
\boldsymbol{\beta^L} & \boldsymbol{\beta^U} & v & e_y & e_{\psi} & \delta& \alpha
\end{bmatrix}^\top,
\]
where we assume the decoupling of the lateral and longitudinal dynamics to sample the vectors \(\boldsymbol {\beta^L}, \boldsymbol {\beta^U}\), with \( \boldsymbol {\beta}^*= \begin{bmatrix}
    \beta^*_{n|k}
\end{bmatrix}_{n\in \mathbb{I}^{k+l}_{k+m}}, *\in\{L,U\}\), and \(\beta^*_{n|k}\) as in \eqref{eq:lane bounds}.
The output slack variables of \(NN_{E^3}^1\) are \( \eta = \begin{bmatrix} \delta_{\underline{a}_y} & \delta_{\bar{a}_y} & \delta_{\underline{j}_y} & \delta_{\bar{j}_y}\end{bmatrix}^\top \).

\begin{figure}[t]
\SetFigLayout{3}{1}
\centering
\subfigure{\includegraphics[width=0.85\columnwidth]{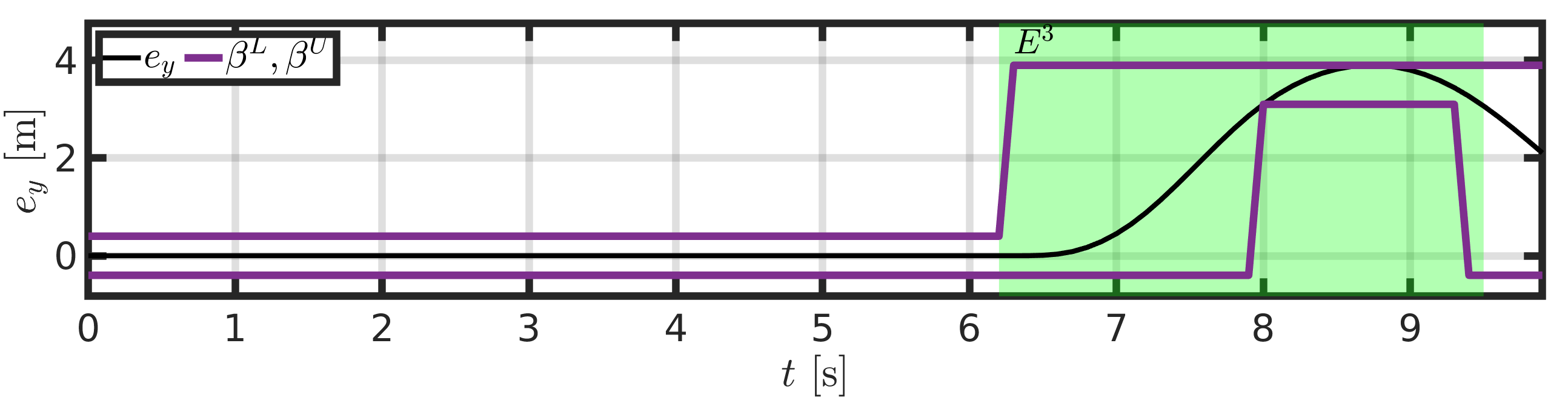}}
\hfill
\centering
\subfigure{\includegraphics[width=0.85\columnwidth]{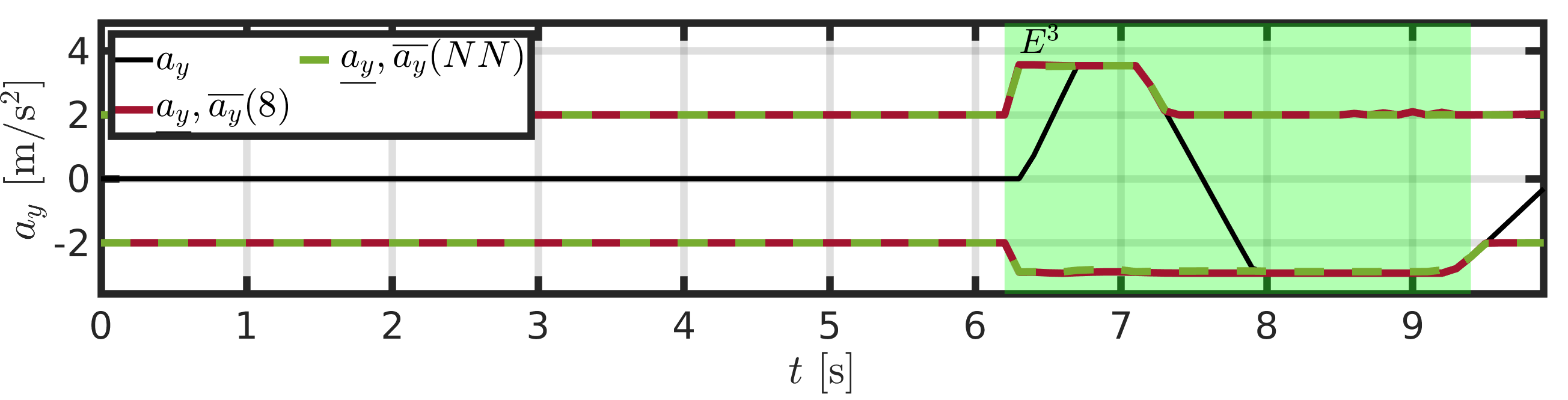}}
\hfill
\centering
\subfigure{\includegraphics[width=0.85\columnwidth]{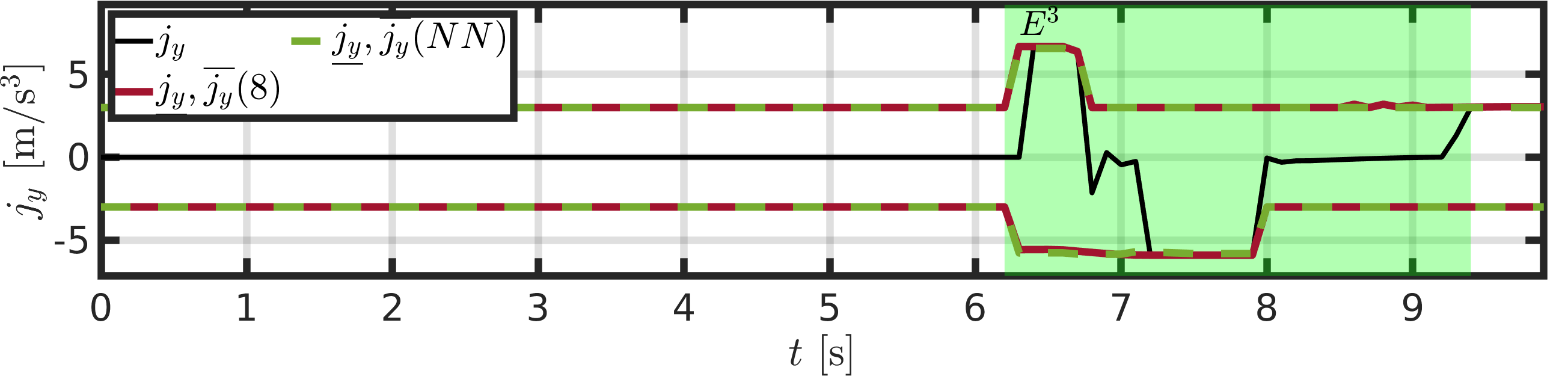}}
\caption{Closed-loop trajectories for Scenario 2: lateral position error (top), lateral acceleration (center), and lateral jerk (bottom).}
\label{fig:LC_traj_ZOD}
\end{figure}

The results obtained by applying Algorithm~\ref{alg:1} are shown in Fig. ~\ref{fig:LC_traj_ZOD}. The green area highlights the time window during which the controller switches to relaxation mode \(E^3\), due to the infeasibility of the relaxation modes \(E^1\) and \(E^2\).
When the surrounding vehicle is predicted to invade the ego lane, the lateral position bounds \(\beta^L,\beta^U\) are updated to initiate an evasive left-lane change maneuver. This results in constraint \eqref{eq:g_lat_safe} violating Assumption~\ref{a:unknown_constraints}. However, the relaxation of the lateral acceleration and jerk comfort bounds enables the controller to preserve feasibility---and thus safety.




\vspace{1em}
\begin{Remark}
While lexicographic MPC formulations can be used to prioritize constraints satisfaction, our approach offers greater flexibility and lower computational complexity. 
Indeed, in the context of prioritized constraint satisfaction considered in this paper, traditional lexicographic optimization requires a defining hard and soft constraints beforehand. This rigid structure can be overly restrictive in dynamic driving scenarios, such as for example, when braking alone is insufficient and a lane change becomes necessary. Our approach, instead, allows defining multiple combinations of hard and soft constraints, each assigned with a priority. 
This allows the controller to adaptively balance objectives based on real-time conditions, avoiding unnecessary conservatism. Furthermore, our formulation avoids the sequential optimization required by standard lexicographic MPC, leading to significantly reduced computational overhead, which is critical for real-time autonomous driving applications.
\end{Remark}

\begin{figure}[t]
\SetFigLayout{2}{1}
\centering
\subfigure{\includegraphics[width=0.9\columnwidth]{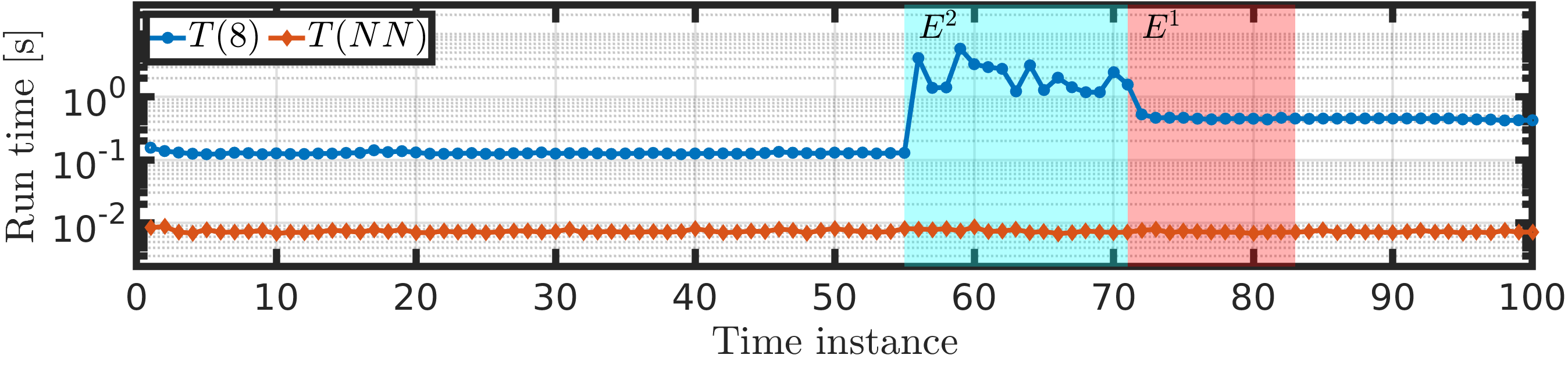}}
\hfill
\centering
\subfigure{\includegraphics[width=0.9\columnwidth]{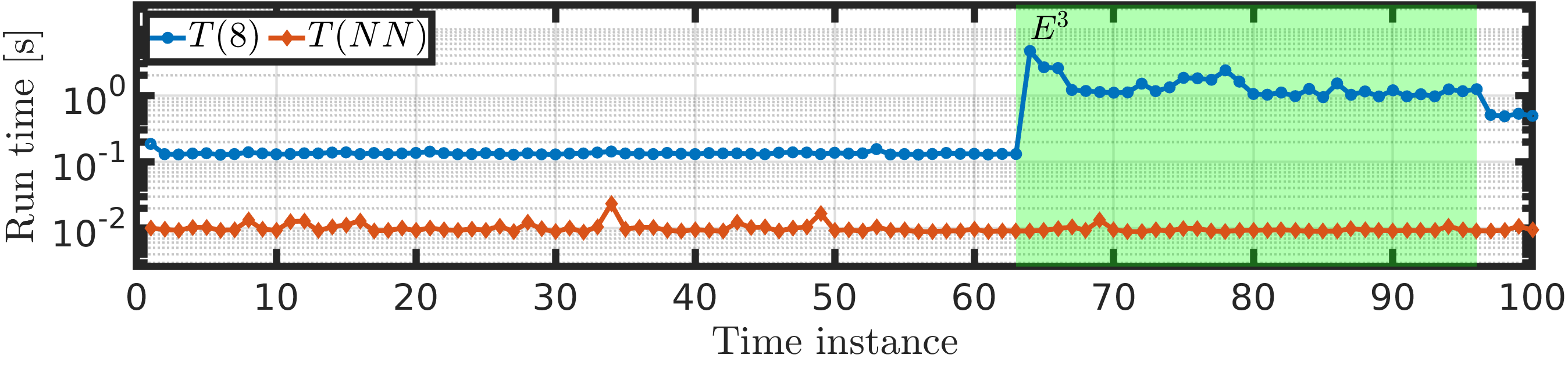}}
\caption{Computation times comparison in Scenario 1 (top) and 2 (bottom) : Problem \eqref{eq:slack_opt} vs. NN.}
\label{fig:comp_times}
\end{figure}

\section{Conclusion}
This paper demonstrates how a Safe MPC scheme can be deployed in automated driving applications to provide the designer with full control of the constraints relaxation. Two SMPC controllers are designed for ACC and automated lane change, respectively, where the designer can straightforwardly decide the constraints that can be relaxed to guarantee the satisfaction of user-defined hard constraints, thus achieving the desired vehicle behavior.

The presented results demonstrate that the typical desired behaviors of ACC and automated lane change systems can be achieved with minimal tuning effort. Furthermore, the run-time overhead introduced by the priority-based relaxation mechanism, is reduced thanks to a learning-based design, thus enabling real-time implementation of the proposed approach.

\section*{Acknowledgments}
We would like to thank Zenseact for providing the data used in this study. Their support was essential to the completion of this research.

\bibliographystyle{IEEEtran}
\bibliography{references}

\begin{thebibliography}{10}
\providecommand{\url}[1]{#1}
\csname url@samestyle\endcsname
\providecommand{\newblock}{\relax}
\providecommand{\bibinfo}[2]{#2}
\providecommand{\BIBentrySTDinterwordspacing}{\spaceskip=0pt\relax}
\providecommand{\BIBentryALTinterwordstretchfactor}{4}
\providecommand{\BIBentryALTinterwordspacing}{\spaceskip=\fontdimen2\font plus
\BIBentryALTinterwordstretchfactor\fontdimen3\font minus \fontdimen4\font\relax}
\providecommand{\BIBforeignlanguage}[2]{{%
\expandafter\ifx\csname l@#1\endcsname\relax
\typeout{** WARNING: IEEEtran.bst: No hyphenation pattern has been}%
\typeout{** loaded for the language `#1'. Using the pattern for}%
\typeout{** the default language instead.}%
\else
\language=\csname l@#1\endcsname
\fi
#2}}
\providecommand{\BIBdecl}{\relax}
\BIBdecl

\bibitem{batkovic2022safe}
I.~Batkovic, M.~Ali, P.~Falcone, and M.~Zanon, ``Safe trajectory tracking in uncertain environments,'' \emph{IEEE Transactions on Automatic Control}, 2022.

\bibitem{batkovic2023experimental}
I.~Batkovic, A.~Gupta, M.~Zanon, and P.~Falcone, ``Experimental validation of safe mpc for autonomous driving in uncertain environments,'' \emph{IEEE Transactions on Control Systems Technology}, 2023.

\bibitem{liu2019recursive}
Z.~Liu and O.~Stursberg, ``Recursive feasibility and stability of mpc with time-varying and uncertain state constraints,'' in \emph{2019 18th European Control Conference (ECC)}.\hskip 1em plus 0.5em minus 0.4em\relax IEEE, 2019, pp. 1766--1771.

\bibitem{kerrigan_soft_2000}
E.~Kerrigan and J.~Maciejowski, ``Soft constraints and exact penalty functions in model predictive control,'' 09 2000.

\bibitem{borrelli_predictive_2017}
F.~Borrelli, A.~Bemporad, and M.~Morari, \emph{Predictive Control for Linear and Hybrid Systems}, 1st~ed.\hskip 1em plus 0.5em minus 0.4em\relax Cambridge University Press.

\bibitem{zheng2011advanced}
T.~Zheng, \emph{Advanced model predictive control}.\hskip 1em plus 0.5em minus 0.4em\relax BoD--Books on Demand, 2011.

\bibitem{kerrigan_designing_2002}
E.~Kerrigan and J.~Maciejowski, ``Designing model predictive controllers with prioritised constraints and objectives,'' in \emph{Proceedings. {IEEE} International Symposium on Computer Aided Control System Design}.\hskip 1em plus 0.5em minus 0.4em\relax {IEEE}, pp. 33--38.

\bibitem{rawlings2017model}
J.~B. Rawlings, D.~Q. Mayne, M.~Diehl \emph{et~al.}, \emph{Model predictive control: theory, computation, and design}.\hskip 1em plus 0.5em minus 0.4em\relax Nob Hill Publishing Madison, WI, 2017, vol.~2.

\bibitem{pauli2021training}
P.~Pauli, A.~Koch, J.~Berberich, P.~Kohler, and F.~Allg{\"o}wer, ``Training robust neural networks using lipschitz bounds,'' \emph{IEEE Control Systems Letters}, vol.~6, pp. 121--126, 2021.

\bibitem{alibeigi2023zenseact}
M.~Alibeigi, W.~Ljungbergh, A.~Tonderski, G.~Hess, A.~Lilja, C.~Lindstr{\"o}m, D.~Motorniuk, J.~Fu, J.~Widahl, and C.~Petersson, ``Zenseact open dataset: A large-scale and diverse multimodal dataset for autonomous driving,'' in \emph{Proceedings of the IEEE/CVF International Conference on Computer Vision}, 2023, pp. 20\,178--20\,188.

\bibitem{andersson2019casadi}
J.~A. Andersson, J.~Gillis, G.~Horn, J.~B. Rawlings, and M.~Diehl, ``Casadi: a software framework for nonlinear optimization and optimal control,'' \emph{Mathematical Programming Computation}, vol.~11, no.~1, pp. 1--36, 2019.

\bibitem{wachter2006implementation}
A.~W{\"a}chter and L.~T. Biegler, ``On the implementation of an interior-point filter line-search algorithm for large-scale nonlinear programming,'' \emph{Mathematical programming}, vol. 106, pp. 25--57, 2006.

\bibitem{jeon2019tracking}
W.~Jeon, A.~Zemouche, and R.~Rajamani, ``Tracking of vehicle motion on highways and urban roads using a nonlinear observer,'' \emph{IEEE/ASME transactions on mechatronics}, vol.~24, no.~2, pp. 644--655, 2019.

\end{thebibliography}

\end{document}